\documentstyle[aps,prl,epsf,floats,amsmath]{revtex} 
\renewcommand\({\left(}
\renewcommand\){\right)}
\renewcommand\[{\left[}
\renewcommand\]{\right]}
\newcommand{\pa}{\partial}
\newcommand{\dd}{{\rm d}}

\newcommand{\be}{\begin{equation}}
\newcommand{\ee}{\end{equation}}
\newcommand{\bea}{\begin{eqnarray}}
\newcommand{\eea}{\end{eqnarray}}

\newcommand{\half}{\frac{1}{2}}

\newcommand\TeV{\,\mbox{TeV}}
\newcommand\GeV{\,\mbox{GeV}}
\newcommand\MeV{\,\mbox{MeV}}

\newcommand\eV{\,\mbox{eV}}

\newcommand\mpl{M_{\rm P}}
\newcommand\Mpl{M_{\rm P}}

\newcommand\bfx{{\bf x}}

\draft

\begin{document}
\twocolumn[\hsize\textwidth\columnwidth\hsize\csname
@twocolumnfalse\endcsname

\title{The curvaton scenario in supersymmetric theories}

\author{Marieke Postma} 
\address{The Abdus Salam International Center for Theoretical Physics, 
Strada Costiera 11, 34100 Trieste, Italy}
\date{November 2002}

\maketitle  
             
\begin{abstract}
We analyze the curvaton scenario in the context of supersymmety.
Supersymmetric theories contain many scalars, and therefore many
curvaton candidates. To obtain a scale invariant perturbation
spectrum, the curvaton mass should be small during inflation $m \ll
H$. This can be achieved by invoking symmetries, which suppress the
soft masses and non-renormalizable terms in the potential. Other
constraints on the curvaton model come from nucleosynthesis, gravitino
overproduction, and thermal evaporation.  The curvaton coupling to
matter should be very small to satisfy these constraints ({\it e.g.} $h
\lesssim 10^{-8}$ for typical soft masses $m \sim \TeV$).
\end{abstract}

\vskip2.0pc]

\renewcommand{\thefootnote}{\fnsymbol{footnote}}
\setcounter{footnote}{0}


\section{Introduction}

It is now widely believed that the early universe went through a
period of superluminal expansion, known as inflation.  In addition to
explaining the isotropy and homogeneity of our observable universe,
inflation can generate the seeds for structure formation.  In the
usual picture the quantum fluctuations of the slowly rolling inflaton
field ``freeze in'' soon after horizon exit, and become essentially a
classical perturbation which remains constant until the moment of
horizon re-entry.  The produced perturbations only depend on the
potential during inflation, and therefore severely constrain possible
inflationary models.

Recently there has been a revived interest in the idea that not the
perturbations in the inflaton field are responsible for the cosmic
microwave background (CMB) fluctuations, but instead isocurvature
fluctuations in another scalar field, the ``curvaton'' field, that is
sub-dominant during inflation. After inflation the isocurvature
fluctuations of the curvaton field are converted into adiabatic ones.
If at decay the contribution of the curvaton to the energy density is
considerable, this can lead to perturbations of the observed size. The
generation of curvature does not depend on the nature of inflation,
beyond the requirement that the Hubble constant must be nearly
constant.  This mechanism of generating perturbations was first
described years ago~\cite{early}, but it did not attract much interest
until recently~\cite{sloth,wands,moroi,curvaton,decouple}.

The usual set up is the following~\footnote[2]
{For an alternative implementation of the curvaton model
see~\cite{mar}}
: During inflation the curvaton field acquires a large expectation
value if it has a negative mass squared, or if its mass is much
smaller than the Hubble constant. In the post-inflationary epoch the
Hubble expansion acts as a friction term in the equations of motion,
and the field remains effectively frozen at a large field value. This
stage ends when the Hubble constant becomes of the order of the
curvaton mass, $H \sim m_\phi$, at which point the curvaton starts
oscillating in the quadratic potential.  The curvaton behaves as cold
matter; its energy density red shifts as $\rho_\phi \propto a^{-3}$,
with $a$ the scale factor of the universe. After inflaton decay, the
universe becomes radiation dominated, with the energy density in
radiation red shifting as $\rho_I \propto a^{-4}$. During this state
of mixed matter, {\it i.e.}, radiation and cold dark matter, the
isocurvature perturbations are transformed into adiabatic ones.  The
perturbations grow until the curvaton becomes to dominate the
universe, or if this never happens, until the curvaton decays.

In this paper we reanalyze the curvaton scenario in the context of
supersymmetry (SUSY). SUSY theories contain many scalar fields, and
therefore many possible curvaton candidates.  However, the finite
energy density in the early universe breaks supersymmetry, and induces
soft masses of the order of the Hubble constant for all scalar
fields~\cite{softmass,flat}. Such heavy fields lead to a perturbation
spectrum with a large spectral tilt, in contradiction with
observations.  Symmetries can be invoked to protect scalars from large
soft masses.  In D-term inflation Hubble induced soft masses are
forbidden by gauge invariance~\cite{dterm}. In no-scale gravity
models, and generalizations thereof, there is a Heisenberg symmetry
that protects scalar fields from soft mass terms at tree
level~\cite{noscale}.  Another possibility is that the curvaton field
is a (pseudo) Goldstone boson.

Non-renormalizable terms in the potential can lift the flat potential
for the large field values needed in the curvaton scenario.  For the
curvaton model to work such terms should be very small or absent up to
some high order. Here SUSY helps, since the non-renormalization
theorem assures that operators absent in the superpotential are absent
at all scales.

There are several constraints on the curvaton model. The produced
perturbations should be adiabatic (which requires that the curvaton
decays after inflaton decay), nearly scale invariant (the curvaton
mass should be much smaller than the Hubble constant during
inflation), and of the correct magnitude.  For the perturbations to be
sizable, the curvaton energy density needs to be a significant
fraction of the total energy density in the universe.  To avoid
gravitino overproduction the reheat temperature should be sufficiently
low.  Curvaton decay should not destroy the nucleosynthesis
predictions. Furthermore, thermal damping and evaporation should be
taken into account.

In the next section we discuss these various constraints on the
curvaton model in more detail. In section \ref{mass} we consider the
possible (soft) mass terms that can arise in SUSY theories, and
analyze the parameter space for the various cases.  We end with
conclusions.

\section{Constraints}

\subsection{Density perturbations}

The density perturbations for the curvaton field have been analyzed
in~\cite{wands}. We briefly review their results.

Consider the curvaton with minimal kinetic term and scalar potential
$V(\phi)$.  We can expand the field in a classical part plus quantum
fluctuations
\be 
\phi(\bfx) = \phi + \delta\phi(\bfx).
\ee
The unperturbed and perturbed curvaton field satisfy respectively
\be 
\ddot\phi +3H_I \dot\phi  + V_\phi = 0 \,, 
\label{eq0} 
\ee 
\be 
\delta \ddot{\phi}_k + 3H_I \delta \dot{\phi}_k + \( ({k}/{a})^2 +
V_{\phi\phi} \) \delta\phi_k = 0 \,. 
\label{eqdelta} 
\ee
Here $\delta \phi_k$ are the Fourier components of $\delta \phi$.  An
overdot denotes $\pa/\pa t$, and a subscript $\phi$ denotes $\partial
/\partial \phi$.  We have made the first order approximation $\delta
(V_\phi) = V_{\phi\phi} \delta \phi$.  The fluctuations of a generic
massive scalar field generated during de Sitter stage are then found
to be on superhorizon scales (where the gradient term in
Eq.~(\ref{eqdelta}) is negligible)
\be 
|\delta \phi_k| \simeq 
\frac{H_*}{2k^3} \( \frac{k}{aH_*}\)^{3/2-\nu_\phi}, 
\ee
where the subscript $*$ denotes the time of horizon exit.  Further,
$\nu_\phi = (9/4 - m/H^2)$. For $\eta_\phi = (m / 3H^2) \ll 1$ one has
$3/2 - \nu_\phi \simeq \eta_\phi$ and the spectrum is nearly scale
invariant. To be more precise, the spectral tilt of the perturbation
is given by
\be
n_\phi \equiv \frac{d\ln {\mathcal P}_\phi }{ d\ln k}
 = 2 \eta_\phi - 2 \epsilon_H,
\label{spectrum} 
\ee 
and it is assumed that $\epsilon_H \equiv \dot{H}/H \ll 1$. 

The field remains overdamped until $H \sim m$ when the field starts
oscillating in the potential. The potential rapidly approaches a
quadratic form, after which the fractional perturbation $\delta \phi /
\phi$ remains constant. We parameterize the change in fractional
density between the end of inflation and the moment the potential
approaches a quadratic form by the parameter $q$:
\be \( \frac{\delta \phi}{\phi} \)_{\!{\rm osc}} 
= q \( \frac{\delta \phi}{\phi}
\)_*,
\label{q}
\ee
where the subscript osc denotes the quantity at the onset of
oscillations, when $H \sim m$. For a quadratic or flat potential the
evolution Eqs.  (\ref{eq0},~\ref{eqdelta}) for $\phi$ and $\delta
\phi$ are the same and $q=1$.

The oscillating curvaton field behaves as non-relativistic
matter. After inflaton decay, the energy density becomes a mixture of
radiation and matter, and isocurvature perturbation are transformed
into adiabatic ones.  This period ends when the curvaton becomes to
dominate the energy density or, if that never happens, when it decays.

The prediction of the curvaton model for the curvature perturbation
can be written as (for small $\eta_\phi$):
\be
{\mathcal P}_\zeta^{1/2} = \frac{r_{\rm dec} q }{3 \pi} 
\frac{H_*}{\phi_*},
\label{cmb}
\ee
where the subscript dec means the corresponding quantity evaluated at
the time of $\phi$-decay, {\it i.e.}, when $H \sim
\Gamma_\phi$. Furthermore, we have defined the parameter $r$ as the
ratio of energy density in the curvaton field to the total energy
density in the universe $ r = \rho_\phi /\rho $.  The COBE data
requires
\bea
{\mathcal P}_\zeta^{1/2} ({\rm COBE}) &= 4.8 \times 10^{-5},
\nonumber \\
n({\rm COBE}) & = 0.93 \pm 0.13.
\label{cobe}
\eea
Moreover, the non-detection of tensor perturbations sets an upper
limit on the energy density during inflation
\be 
H_* \lesssim 10^{14} \GeV.
\label{hmax}
\ee

\subsection{Initial conditions}

The initial field value of the curvaton at the beginning of inflation
is a free parameter.  The only constraint is that the energy density
in the curvaton field is sub-dominant.  The inflaton energy density is
$\rho_I \sim H_I^2 \mpl^2$, with $H_I$ the Hubble constant during
inflation. This means that for masses $m^2 \ll H_I^2$ a typical
initial field value will be large $\phi \gg \mpl$.  

We consider scalar potentials of the form
\be
V(\phi) = \half m^2 \phi^2 + \frac{\lambda}{M^n} \phi^{4+n}.
\label{pot}
\ee
The non-renormalizable operators, suppressed by the Planck scale or
some other ultraviolet cutoff, will generate an effective mass $\delta
m_{\rm HO}^2 = \pa^2 V / \pa \phi^2 \equiv V_{\phi \phi} > H$ for
large initial field values.  The curvaton is underdamped, and
decreases exponentially fast.

The effective mass squared term during inflation can be positive or
negative.  The finite energy density in the early universe breaks
supersymmetry. This leads to soft mass terms of the form $m^2 = c
H_I^2$.  The sign of $c$ is determined by the K\"ahler potential.

If the effective mass squared is negative and the initial curvature
amplitude is large, the curvaton field will approach its minimum
exponentially fast.  If on the other hand, the curvaton is initially
at the origin, or at some small field value, the field is highly
damped. It will approach the minimum at a rate $\phi \propto m^2 t$
(for constant $m$). At $t\sim H_I/m^2$ the motion goes non-linear and
the classical field reaches the minimum exponentially fast.  We will
assume that inflation last sufficiently long for the field to be in
its minimum.\footnote[3] {Large initial fields decrease exponentially
to the minimum (or if it overshoots) or lower field values. Therefore,
it requires extreme (exponentially) fine tuning to obtain field values
$\phi > \phi_{\rm min}$ at the end of inflation.  It is possible to
obtain $\phi < \phi_{\rm min}$, if inflation ends before the classical
motion for small amplitudes goes non-linear.  But this will not
increase parameter space.}

If the mass squared is positive during inflation, and the field has
large initial values, it will decrease exponentially fast until $m
\lesssim H_I$, and the field becomes overdamped.  The field value
typically freezes at $\phi \sim H_I$.  After that, the field decreases
only linearly until the classical motion becomes non-linear at $t\sim
H_I/m^2$. Quantum fluctuations grow during inflation as $\langle
\phi^2 \rangle \approx H_I^3 t / 4 \pi^2$ until they reach at $t \sim
H_I/m^2$ a saturation value~\cite{qf}
\be
\langle \phi^2 \rangle \approx \frac{3}{8 \pi^2} 
\frac{H_I^4}{m^2},
\label{qf}
\ee
with coherence length (provided the classical field has a small
amplitude, $\phi \ll H_I/m$)
\be
l \sim H_I^{-1} \exp \( \frac{ 3 H_I^2}{2m^2} \). 
\label{coh}
\ee
The low momentum modes of these fluctuations will be indistinguishable
from a classical field with an amplitude $\phi_I \approx \sqrt{\langle
\phi^2 \rangle}$. For the field to be homogeneous on the current
horizon size $m^2/H_I^2 < 1/40$.  If higher order terms in the
potential are non-negligible, $m^2$ should be replaced by $\delta
m_{\rm HO}^2 = V_{\phi \phi}$ in the above formulas.

\subsection{Domination}

When $H \sim m$ the curvaton field start oscillating in its potential;
it behaves as cold dark matter with $\rho_\phi \propto a^{-3}$.  After
inflaton decay the universe becomes radiation dominated, its energy
density red shifting as $a^{-4}$.  The fractional energy density in
the curvaton field increases until a maximum value at curvaton decay:
\be r_{\rm dec} = \( \frac{\rho_\phi}{\rho} \)_{\rm dec}
\sim \frac{1}{h} \frac{\phi_{\rm osc}^2}{\Mpl^2} \( \frac{\Gamma_I}{m}
\)^\alpha,
\label{r_dec}
\ee
with $\alpha = 0$ for $\Gamma_I > m$, and $\alpha = 1/2$ for $\Gamma_I
< m$.  We have assumed instant reheating, with reheating temperature
$T \sim g_*^{1/4} \sqrt{\Mpl \Gamma_I}$. Above $T \sim 10^{-1} \GeV$
the effective degrees of freedom $g_* = {\mathcal O}(100)$, whereas
round MeV temperatures it drops to $g_* = {\mathcal O}(10)$.  Further,
we used $\Gamma_\phi \sim h^2 m$, with $h$ the coupling, yukawa or
gauge, of the curvaton to matter.  The curvaton field dominates the
energy density in the universe at decay for $r \gtrsim 1/2$.

As can be seen from Eq.~(\ref{r_dec}), to avoid a period of curvaton
dominated inflation, the curvaton cannot dominate the energy density
before the onset of oscillation. This translates into
\be
\phi_{\rm osc}  < \mpl.
\label{phi_inflation}
\ee

\subsection{Thermal constraints}

The curvaton field must decay before $T \sim \MeV$, to not destroy the
successful predictions of nucleosynthesis. This gives
\be
\Gamma_\phi \sim h^2 m_\phi \gtrsim 10^{-25} \GeV.
\label{bbn}
\ee

The reheat temperature of the universe after inflaton decay cannot be
too high; to avoid gravitino overproduction requires $T_I < 10^6 -
10^9 \GeV$.  However, gravitinos produced by inflaton decay are
diluted by the entropy generated in the curvaton decay.  If the
curvaton has a sub-dominant energy density at decay the dilution is
negligible. The dilution factor $\Delta \equiv s_{\rm after}/ s_{\rm
before}$ is
\be 
\Delta  = 
\frac{1}{h} \frac{\phi_{\rm osc}^2}{\mpl^2} 
\( \frac{\Gamma_I}{m} \)^\alpha
\label{dilution}
\ee
This implies $T_I < \Delta (10^6 -10^9) \GeV$, or
\be
\Gamma_\phi \lesssim 
(10^{-7} - 10^{-1}) \GeV \( \frac{\phi_{\rm osc}}{\mpl} \)^{4}
\( \frac{m}{\Gamma_I} \)^{1-2\alpha}. 
\label{gravitino}
\ee
Similarly, any preexisting baryon asymmetry gets diluted by the
entropy production.  Therefore, if $\Delta > 10^{10}$, the baryon
asymmetry must be generated in the out-of-equilibrium curvaton decay,
or at lower temperatures.

Finite temperature effects can lead to thermal damping or early
thermal decay of the condensate \cite{AD,AD_thermal}.  Large thermal
masses $\delta M_{th} \gtrsim H > m$ may induce early oscillations,
which reduces the energy stored in the condensate.  When the
temperature is higher than the effective mass of the particles coupled
to the condensate, {\it i.e.}, $T \gtrsim h \phi$, the curvaton
receives a thermal mass $\delta M_{\rm th}^2 = (1/4) h^2 T^2$. For
lower temperatures $T \lesssim h \phi$ the running of the yukawas is
modified leading to an effective thermal mass $M_{\rm th} = h^4 T^4 /
\phi^2$.  Thermal damping is unimportant when
\be 
\frac{\delta M_{\rm th}^2}{H^2} \lesssim 1 \quad {\rm for} \;\; H > m
\label{mass_th}
\ee
It is important to note that even before $H \sim \Gamma_I$ there is a
dilute plasma with temperature $T \sim (T_I^2 \Mpl
H)^{1/4}$~\cite{kt}.

Another effect to be considered is thermal evaporation by particles
which are in equilibrium with the thermal bath, that is for which $
m_{\rm eff} \sim h \phi < T$ \cite{AD_thermal}. For a coupling in the
scalar potenital $V \ni h \phi^2 \chi^2$ the cross section for $\phi
\chi$-scattering is $\sigma \sim h^2 /E_{\rm cm}^2$.  The typical
center of mass energy is $E_{\rm cm} \sim \sqrt{T m}$, the mean of the
typical $\chi$-energy ($\sim T$) and $\phi$-energy ($\sim m$). The
thermal decay rate is $\Gamma_{\rm th} \sim \sigma T^3 \sim h^2 T^2
/m$.  If the condensate evaporates thermally when $H \gtrsim
\Gamma_I$, the isocurvature curvaton are not converted into adiabatic
perturbations, and the curvaton scenario does not work. This happens
when
\be
\Gamma_{\rm th} \gtrsim H  \quad \& \quad
T > h \phi.
\label{gamma_th}
\ee

All thermal constraints can be circumvented if the inflaton sector and
curvaton sector decouple completely, as is proposed in
reference~\cite{decouple}.  In their scenario the inflaton is a hidden
sector field which decays into hidden sector radiation, while the
curvaton field is responsible for reheating of the MSSM sector. Then,
before curvaton decay there is no thermal bath of MSSM particles, and
thermal effects are negligibly small.

\section{Various models}
\label{mass}

The finite energy density in the early universe breaks supersymmetry,
leading to soft masses in the scalar potential which are generically
of the order $H$~\cite{softmass}. Soft mass terms are both induced by
non-renormalizable as by supergravity corrections. In global SUSY,
non-renormalizeble terms in the K\"ahler potential of the form $\delta K
= \int \dd^4\theta \, \Mpl^{-2} \phi^\dagger \phi I^\dagger I$ lead to
mass terms $\delta m^2 \sim (\rho/\Mpl^2) \sim H^2$.  Supergravity
corrections to the Lagrangian likewise induce mass terms of the order
$H^2$.  The mass squared can be either positive or negative, depending
on the specific K\"ahler potential.

However, the curvaton scenario cannot work for $|\delta m| = {\mathcal
O}(H)$, since this would give large deviations from scale dependence
in the density fluctuations, in conflict with observations.  The COBE
requirement Eqs.~(\ref{spectrum},~\ref{cobe}) translates into $ m /
H_I < 10$.  But the constraint can be made stronger.  For a positive
mass squared a large coherence length for the fluctuations is needed,
which requires $ m / H_I < 1/40$.  For a negative mass squared, there
is another bound.  After horizon exit the zero mode and fluctuations
evolve according to Eqs.~(\ref{eq0},~\ref{eqdelta}).  Since the mass
is negative $m^2 = - c H^2$, the classical field is at its minimum.
Substituting this in the equation for the fluctuations yields a
positive effective mass squared $m_{\rm eff}^2 = [(n+4)(n+3)-1] c
H^2$.  When the effective mass is of the order $H$ damping is
efficient, and the fluctuations will decrease exponentially fast
during the last 60 e-folds of inflation.  This gives a bound $m / H_I
\lesssim 1/60$.

Therefore, for the curvaton scenario to work within the context of
supersymmetry we need to invoke symmetries to protect the scalar field
from obtaining large soft masses.

When inflation is driven by $D$-terms, gauge invariance forbids the
appearance of soft masses for the scalar fields~\cite{dterm}.  The
scalar mass are set by the low energy SUSY breaking, and are typically
of the order $\sim m_{3/2}$.  At the end of inflation, the vacuum
energy density stored in $D$-terms is transferred to kinetic and
$F$-terms, which do induce mass terms of the $\delta m^2 \sim H^2$.
The sign of the mass term depends on the K\"ahler potential.

No-scale type gravity models possess an extra, so-called Heisenberg
symmetry, which forbids soft mass terms at tree level~\cite{noscale}.
Masses are induced radiatively, and are suppressed by loop factors
\be 
\delta m^2 \sim \frac{c}{4\pi^2} h^2 H^2, 
\label{mass_ns}
\ee
with $c = {\mathcal O}(1)$, which can be positive or negative
depending on whether gauge or yukawa coupling dominate, and on which
field (hidden sector, matter field, dilaton) is the inflaton.  For
$|\delta m^2| > m^2$, where the curvaton mass is typically set by low
energy SUSY breaking, the mass squared can be negative during
inflation.  Since the gravitino mass decouples in no-scale gravity,
there is no problem related to gravitino overproduction.

Pseudo-Goldstone bosons are protected from soft corrections; the mass
is set by the breaking scale of the global symmetry.  The analysis for
this case is, apart from constraints concerning gravitino production,
the same as for standard model curvaton fields.

\subsection{Negative mass squared --- no-scale inflation}
\label{s_ns1}

In this section we consider the parameter space for curvaton fields
with no-scale type masses of the form
\be
\delta m^2 \sim - 10^{-2} h^2 H^2.  
\ee
For the induced mass to be the dominant term, we also require that
during inflation the effective mass is negative, {\it i.e.},
\be
m^2 < |\delta m^2|
\label{small_m}
\ee
Then during inflation the $\phi$-field is driven to its classical
minimum
\be
\phi_{\rm min} \sim (H^2 M^n/\lambda)^{1/(n+2)}.
\label{min}
\ee
In the post-inflationary evolution the field keeps tracking its
instantaneous minimum until $|\delta m^2| \sim m^2$, at which point
the mass becomes positive, and the field freezes until $H \sim m$. The
higher order terms in the potential Eq.~(\ref{pot}) are of the same
order as the mass term at the minimum, and therefore do not alter the
conclusion that the fluctuation spectrum Eq.~(\ref{spectrum}) is flat.
However, the higher order terms are important for the evolution of the
fluctuations. With the classical field at its minimum, it follows from
Eq.~(\ref{eqdelta}) that the fluctuations have a positive effective
mass squared $m_{\rm fluc}^2 = + \[ (n+4)(n+3)-1 \] 10^{-2} h^2
H^2$. While the zero mode decreases for $|\delta m| > m$, fluctuations
are overdamped and remain frozen. This leads to a $q$-factor
Eq.~(\ref{q})
\be 
q = \frac{\phi_*}{\phi_{\rm osc}},
\ee
with $\phi_{\rm osc}$ the classical minimum Eq.~(\ref{min}) at $|\delta
m| \approx 0.1 h H \sim m$.  The fraction of energy density stored in
the curvaton field at decay is 
\be
r_{\rm dec} = 
\frac{1}{h}
\frac{\phi_{\rm osc}^2}{\Mpl^2} 
\( \frac{\Gamma_I}{m} \)^\alpha.
\ee
The CMB constraint Eqs.~(\ref{cmb},~\ref{cobe}) then reads
\be
\frac{H_*}{h} \( \frac{m^2 M^n}{\lambda} \)^{\frac{1}{n+2}} 
\( \frac{\Gamma_I}{m} \)^\alpha
= 3 \times 10^{33} \GeV^2.
\ee
Domination occurs for $r \sim 1$, or
\be
h \lesssim \( \frac{m}{\mpl} \)^{4/(n+2)}
\ee

The curvaton scenario does not work with at quartic term in the
potential. The CMB constraint for $n=0$, together with bound on the
Hubble constant during inflation Eq.~(\ref{hmax}) and nucleosynthesis
constraints Eq.~(\ref{bbn}) eliminates all parameter space.

\begin{figure}[t]
\centering
\hspace*{-5.5mm}
\leavevmode\epsfysize=6cm \epsfbox{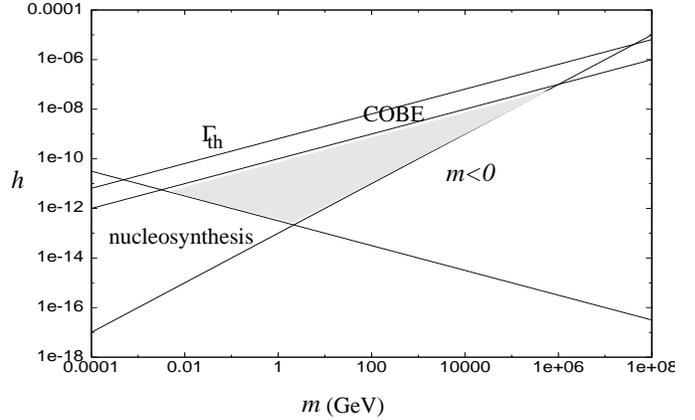}\\
\caption[fig.1] {Parameter space for the curvaton model in no-scale
type gravities with a soft mass term $\delta m^2 \approx - 10^{-2} h^2
H^2$ during inflation, for the parameters $M/\sqrt{\lambda} =\mpl'$,
$n=2$, and $\Gamma_I >m$. The curvaton scenario works for masses and
couplings in the shaded area.}
\end{figure}

If the potential if lifted by a $\phi^6$-term, the CMB constraint
becomes
\be
h < 10^{-8} \( \frac{H}{10^{14} \GeV} \) 
\( \frac{m}{10^{4} \GeV} 
\frac{M/ \sqrt{\lambda}}{\mpl} \)^{1/2}
\( \frac{\Gamma_I}{m} \)^{\!-\alpha}
\ee
This bound is shown in Fig. 1 for $M /\sqrt{\lambda} =\Mpl$ and
$\Gamma_I > m$ ($\alpha = 0$), together with the constrains coming
from nucleosynthesis, the requirement of having a negative effective
mass during inflation Eq.~(\ref{small_m}), and from thermal
evaporation Eq.~(\ref{gamma_th}).  

If the condensate evaporates thermally when $H \gtrsim \Gamma_I$, no
adiabatic perturbations are generated, and the curvaton scenario
fails. The bound is strongest in the thermal plasma after inflaton
decay, for $H \lesssim \Gamma_I$. From Eq.~(\ref{gamma_th}) it then
follows that thermal evaporation occurs for couplings in the range
\be 
10^{-9} \( \frac{m}{\GeV} \)^{1/2}  
< h < 
1 \( \frac{\mpl}{M/\sqrt{\lambda}} \)^{1/2}, 
\ee
In the same coupling range damping by thermal masses can become
important.

There are no constraints from gravitino production, as the gravitino
mass can be arbitrary high in no-scale models. Further, $\phi_{\rm
osc} < \mpl$ automatically, and therefore $\phi$-driven inflation does
not occur Eq.~(\ref{phi_inflation})

The $\phi$-field can only act as the curvaton for small masses $m
\lesssim 10^5 \GeV$ and small couplings $h \lesssim 10^{-8}$.  For
typical soft masses $m \sim \TeV$, one needs to fine tune all
parameters $H \to 10^{14} \GeV$, $|\delta m| \to m$, $M/\sqrt{\lambda}
> \Mpl$ and $h \to 10^{-7}-10^{-8}$ for the curvaton scenario to work.

For higher order operators $n > 2$ the CMB constraint allows for
larger values of the coupling.  However, not much parameter space
opens up as $h$ is bounded from above to avoid early thermal
evaporation.

\subsection{Positive mass squared --- no-scale inflation and Goldstone bosons}
\label{s_ns2}

In this section we discuss the parameter space for curvaton models
with a positive mass squared $0 < m^2 \ll H_I^2$.  Such mass terms can
arise in no-scale supergravity, or alternatively the curvaton can be a
Goldstone boson.  In $D$-term inflation the mass is protected from
soft corrections during inflation, but not afterwards; we will discuss
this case in the section~\ref{dterm}.

For the moment we ignore possible non-renormalizable operators, but we
will discuss them in the next subsection. Furthermore, we approximate
the Hubble constant as constant during inflation, {\it i.e.}, we set
$H_* \approx H_I$.  Here $H_*$ is the Hubble constant when density
fluctuations of the size of our present horizon leave the horizon,
which happens some 60 e-folds before the end of inflation.  $H_I$ is
the Hubble constant at the end of inflation, which sets the magnitude
of the condensate.

During inflation a condensate is formed with magnitude given by
Eq.~(\ref{qf}). The curvature perturbation is given by Eq.~(\ref{cmb})
with $q \approx 1$. If the curvaton comes to dominate the universe, $r
\approx 1$ (see Eq.~(\ref{r_dec})), and the COBE constraint becomes
\bea
h &\lesssim& 10^{-22} \( \frac{m^2}{\GeV^2} \) 
\( \frac{\Gamma_I}{m} \)^{\alpha} 
\nonumber \\
m &\approx& 10^{-4}H_I.
\label{ns2_dom}
\eea
For larger values of the coupling, the curvaton energy density remains
sum-dominant. Then $r < 1$ and the COBE constraint reads
\bea
&&h = 10^{-34} \( \frac{H_I^3}{m \GeV^2} \) 
\( \frac{\Gamma_I}{m} \)^{\alpha},   
\nonumber \\[.2cm]
&&10^4 m \lesssim H_I \lesssim 10^{14} \GeV.
\label{ns2_non_dom}
\eea
For couplings in the range $10^{-9} \sqrt{m /\GeV} < h < h_{\rm up} $
with 
\be 
h_{\rm up} \sim \left\{
\begin{array}{*{2}{l}}
10^2 \( \frac{\Gamma_I H \GeV^2}{m^4} \)^{1/4} 
\( \frac{\Gamma_I}{H} \)^{\beta/2} 
& \quad (r \sim 1)\\
10^{-8} \( \frac{H^3 m^2}{\GeV^5} \)^{1/10} 
\( \frac{\Gamma_I}{m} \)^{2 \alpha/5}
\( \frac{\Gamma_I}{H} \)^{3\beta/20} 
& \quad (r < 1)\\
\end{array}
\right.  
\ee
the condensate is destroyed by thermal scattering before the adiabatic
perturbations are generated.  Here $\beta = 1$ for $H > \Gamma_I$, and
$\beta =0$ for $H < \Gamma_I$. These constraints are plotted in Fig.~2
for the case $\Gamma_I > m$. Also shown are the bounds from
nucleosynthesis and from $\phi$-dominated inflation
Eq.~(\ref{phi_inflation}). If non-renormalizable terms lift the flat
direction at a cutoff scale $\Lambda \lesssim \mpl$, the constraint
$\phi_{\rm osc} \lesssim \mpl$ is automatically satisfied, and the
bound from $\phi$-dominated inflation should be re-interpreted as a
cutoff on the validity of the theory: for higher couplings
non-renormalizable terms can no longer be neglected.

\begin{figure}[t]
\centering
\hspace*{-5.5mm}
\leavevmode\epsfysize=6cm \epsfbox{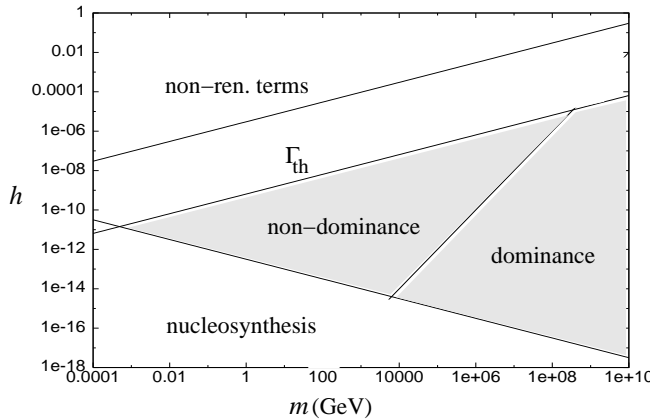}\\[3mm]
\caption[fig.2] {Parameter space for models with a positive mass
squared during inflation, for $\Gamma_I >m$. The curvaton scenario
works for masses and couplings in the shaded area.}
\end{figure}

In the allowed region of parameter space in Fig.~2, the Hubble-induced
mass term in no-scale gravities is smaller than the low energy mass:
$|\delta m| \sim 10^{-1} h H_I < m$. Hence, the exclusion plot based
on a constant mass is also applicable to this class of models.  As
mentioned before, there are no constraints from gravitino
overproduction in no-scale gravity, as the gravitino mass can be
arbitrary high.

Gravitino production does constrain the Goldstone boson curvaton
field. The entropy production from the late-time curvaton decay is
suppressed by the small factor $(\phi_{\rm osc}/\mpl)^4$ (see
Eq.~{\ref{dilution}). Only for masses close to the upper bound $m \to
10^{-4} H_I \sim 10^{10} \GeV$ can it be significant.  For smaller
masses the dilution is negligible small, and the only way to avoid
gravitino overproduction is to have $T_I < 10^6 -10^9 \GeV$. This
means $\Gamma_I < m\:$ ($\alpha =1/2$).  Smaller decay rates are
needed to get curvaton domination. Furthermore, the constraints from
$\phi$-dominated inflation, and from the COBE data
Eq.~(\ref{ns2_non_dom}) become tighter. This last constraint is
equivalent to requiring the curvaton to decay after inflaton
decay. The results are plotted in Fig.~3 for $T_I = 10^7 \GeV$.

\begin{figure}[t]
\centering
\hspace*{-5.5mm}
\leavevmode\epsfysize=6cm \epsfbox{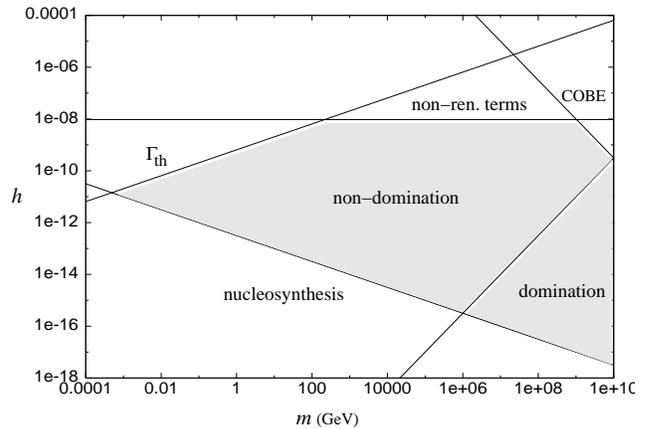}\\[3mm]
\caption[fig.3] {Constraints from gravitino production for $T_I =10^7
\GeV$}
\end{figure}

\subsection{Positive mass squared --- non-renormalizable terms}
\label{s_ho}

The non-renormalizable terms in the potential Eq.~({\ref{pot}) can
alter the picture described in the previous subsection.  Higher order
terms are negligible when $V_{\phi \phi} < m^2$ during inflation, that
is when
\be
m < \left[ 
\sqrt{10^{16} \frac{3}{8 \pi^2}} \(\frac{H_*/m}{10^4}\) 
\right]^{(n+2)/2n}.
\ee
In the case of curvature domination the COBE data requires $H \sim
10^4 m$.  Taking $M^n/\lambda = \mpl$, it follows that higher order
terms are negligible only for small enough masses: $m \lesssim 10^3
\GeV$ for $n=2$, $m \lesssim 10^7 \GeV$ for $n=4$, and $m \lesssim
10^8 \GeV$ for $n=6$.  For larger masses the effective mass $m_{\rm
eff}^2 \equiv V_{\phi \phi}$ sets the initial conditions during
inflation.  As long as $m_{\rm eff}^2 \ll H_I^2$ the perturbation are
nearly scale invariant and the curvaton scenario is possible.  We will
analyze this possibility in this subsection.

The effective mass $V_{\phi\phi}$ sets the initial value of the
curvaton, which from Eq.~(\ref{qf}) is
\be
\phi_I \sim 
\(
\frac{3}{8 \pi^2}  
\frac {H^4 M^n}{(n+4)(n+3)\lambda}
\)^{\frac{1}{n+4}}.
\ee
When $H \sim m_{\rm eff}$ the curvaton field starts oscillating.
After a few oscillation the potential approaches a quadratic form. We
approximate $q \approx 1$.  During the first initial oscillation the
curvaton energy density changes as
\be
\rho(H) = \frac{m(H)}{m_(H_I)} \( \frac{a_0}{a} \)^3 \rho(H_I),
\label{rho.ev}
\ee
where $a$ is the scale factor of the universe. The field amplitude at
the onset of quadratic oscillations, when $m(H) = m$, is $\phi_{\rm
osc}^2 = \kappa^{1/2 + \alpha} \phi_I^2$. Here we have parametrized $m
= \kappa \sqrt{V_{\phi \phi}}$.

The COBE constraints Eq.~(\ref{cobe}) for domination $r \sim 1$
translate into
\be \left\{
\begin{array}{lll}
m < 10^4 \GeV C_m,
\quad h < 10^{-16} C_h 
\qquad (n=2)\\
 m < 10^7 \GeV C_m,
\quad h < 10^{-8}\; C_h 
\qquad (n=4)\\
 m < 10^8 \GeV C_m,
\quad h < 10^{-6}\; C_h 
\qquad (n=6)\\
\end{array}
\right.
\label{hod}
\ee
With $C_m = \kappa (M \lambda^{-1/n}/\mpl)$ and $C_h =
\kappa^{1/2+\alpha} (\Gamma_I/m)^\alpha$.  For the non-domination case
$r<1$, the COBE data requires
\be h < \left\{
\begin{array}{lll}
 10^{-20} m^{5/4} \kappa^{-3/4 + \alpha} \(\frac{\Gamma_I}{m}\)^{\alpha} 
\(\frac{M/\lambda^{1/n}}{\mpl}\)^{3/4}
& \quad (n=2)\\
 10^{-15} m \kappa^{-1/2 + \alpha} \(\frac{\Gamma_I}{m}\)^{\alpha} 
\(\frac{M/\lambda^{1/n}}{\mpl}\)
& \quad (n=4)\\
 10^{-13} m^{7/8} \kappa^{-3/8 + \alpha} \(\frac{\Gamma_I}{m}\)^{\alpha} 
\(\frac{M/\lambda^{1/n}}{\mpl}\)^{9/8}
& \quad (n=6)
\end{array}
\right.
\label{hond}
\ee
And $H_I$ is constraint $10^4 m / \kappa \lesssim H_I \lesssim 10^{14}
\GeV$.

The constraints are shown in Fig.~4 for $M^n/\lambda =\mpl^n$,
$\Gamma_I > m$, and $\kappa \to 1$.  Note that one can read off from
the plot the parameter space where $m^2 > V_{\phi\phi}$ during
inflation and higher order terms can be neglected, namely the region
to the left of the COBE constraints. In most of the curvaton dominated
region in Fig.~2 the non-normalizable terms cannot be
neglected. Curvaton domination can only occur if the
non-renormalizable terms are absent to some very high order, or if the
effective mass during inflation is set by the higher order terms.
This conclusion does not change when $\Gamma_I < m$, since both the
COBE constraints and the requirement $r\sim 1$ have the same
$\Gamma_I$ dependence.

For small masses, $m = \kappa \sqrt{V_{\phi\phi}}$ with $k \ll 1$,
Fig.~4 indicates the couplings $h$ needed for the curvaton scenario
after shifting the plot by some power of $\kappa$ (both to lower $h$
and $m$ values, see Eqs.~(\ref{hod},\ref{hond})). Taking $\Gamma_I <
m$, or $M^n/\lambda < \mpl^n$ has a similar effects, the COBE bounds
shifts to lower couplings and smaller masses.

To summarize, the non-renormalizable terms can be neglected only for
sufficiently small masses. But in a large part of parameter space,
where the curvaton dominates the energy density, their contribution to
the mass term during inflation is dominant, unless these operators are
absent to a very high order.  If the higher order terms dominate, the
curvaton scenario may still work, but smaller couplings and masses are
needed.

\begin{figure}[t]
\centering
\hspace*{-5.5mm}
\leavevmode\epsfysize=5.5cm \epsfbox{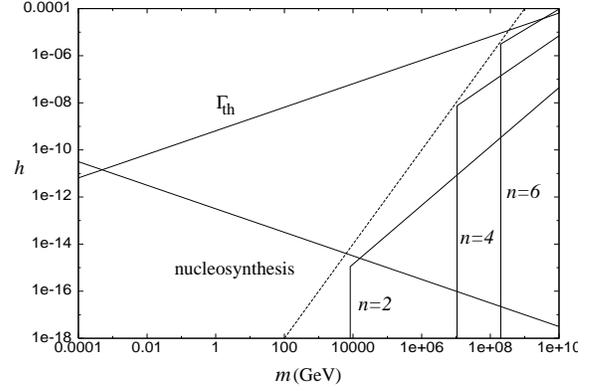}\\[3mm]
\caption[fig.4] {Parameter space when the effective mass is set by
higher order terms $\phi^6$ ($n=2$), $\phi^8$ ($n=4$), or $\phi^{10}$
($n=4$). Here $M = \mpl$, $\lambda =1$, $\Gamma_I > m$, and $\kappa
\to 1$. The dashed line corresponds to the $r=1$ line in the absence
of higher order terms (from Fig.~2), which is added for comparison.}
\end{figure}

\subsection{D-term inflation}
\label{dterm}

In $D$-term inflation there is no Hubble induced soft mass term during
inflation.  However at the end of inflation, the energy in $D$-terms
is transferred to $F$-terms and kinetic energy and a mass term $m^2 =
c H^2$, with $c = {\mathcal O}(1)$ is induced. 

If $c<1$ the field approaches its minimum Eq.~(\ref{min}). The only
difference with the no-scale model discussed in section~\ref{s_ns1} is
the initial field value $\phi_*$. If we ignore the change in $\delta
\phi/\phi$ from turning on the Hubble induced mass, then $\phi_*$
drops out of the COBE constraint. Therefore the parameter space is
given by Fig.~1, with the important difference that the constraint
$m_{\rm eff} < 0$ of Eq.~(\ref{small_m}) does not apply. 

For $c>1$ the field value decreases at the end of inflation, according
to Eq.~(\ref{rho.ev}), and $\phi_{\rm osc} \sim (m/H_{\rm end})^3
\phi_*$ (for $\Gamma_I<m$) with $H_{\rm end}$ the Hubble value at the
end of inflation. Ignoring the change in $\delta \phi/\phi$ from
turning on the Hubble induced mass, the only difference with sections
\ref{s_ho} and \ref{s_ns2} is that the coupling in the COBE
constraints Eqs~(\ref{ns2_dom},\ref{ns2_non_dom}) and
Eqs~(\ref{hod},\ref{hond})) should be replaced by $h \to h (m/H_{\rm
end})^3$.  This decreases parameter space, and moves the domination
curve to the right.

\section{Conclusions}

The finite energy density during inflation breaks supersymmetry, and
induces soft masses of the order of the Hubble constant for all
scalars.  The inflationary perturbation spectrum for these fields is
highly non-Gaussian, and therefore they cannot play the role of the
curvaton.  To avoid this conclusion one has to invoke symmetries, to
keep the soft mass terms small during inflation.

In no-scale type gravities, scalar masses are induced radiatively and
are suppressed to the soft breaking scale ($M_s \sim H$) by loop
factors. These theories have the additional advantage that the
gravitino mass can be arbitrarily large, thereby avoiding problems
with gravitino overproduction. In $D$-term inflation gauge symmetry
forbids soft Hubble-induced masses. At the end of inflation, when
$D$-terms are converted to $F$ terms and kinetic energy, a soft mass
term $|\delta m^2| \sim H^2$ appears.  Pseudo-Goldstone bosons are
protected from soft mass terms. Their mass is set by the amount of
symmetry breaking, and can be small.  Apart from constraints
considering gravitino production, the analysis for this case also
applies to standard model fields.

In all cases the parameter space for the curvaton scenario is
constrained by bounds from nucleosynthesis and early thermal
evaporation.  To avoid thermal evaporation the coupling to matter
should be small. In particular, for typical low energy soft masses $m
\sim \TeV$ and an interaction term in the scalar potential of the form
$V \ni h \phi^2 \chi^2$, thermal evaporation is negligible for $h
\lesssim 10^{-8}$. But we note that thermal constraints can be
circumvented if the inflaton and curvaton sector decouple
completely~\cite{decouple}.

Strong constraints also come from the existence of non-renormalizable
operators in the potential, which are suppressed by some cutoff scale
$M$.  For operators of the form $ V \ni \phi^{n+4}/\mpl^n$, the
curvaton mass is constraint to $m \lesssim 10^3 \GeV$ for $n=2$, $m
\lesssim 10^7 \GeV$ for $n=4$, and $m \lesssim 10^8 \GeV$ for $n=6$.
These bounds are absent for $D$-term inflation, with a negative Hubble
induced mass squared after inflation.

One can ask whether there are any natural candidates for the curvaton
field.  Affleck-Dine condensates have masses of the low energy SUSY
breaking scale, $m \sim \TeV$. Along most flat directions the $\phi^6$
term is absent. However, couplings to matter are too large, and do not
satisfy the curvaton constraint $h < 10^{-8}$.

Moduli and polyoni fields have Planck mass suppressed couplings and
typically decay after nucleosynthesis.  This can be avoided if the
mass term is sufficiently high, $m \gtrsim 100 \TeV$.  In no-scale
gravities this is naturally achieved by giving them a mass of the
order of the gravitino mass.  But that means that also their mass
during inflation is too large ($\delta m^2 \sim H^2$), and they cannot
be the curvaton.

Right-handed sneutrinos can have small couplings to matter. In the sea
saw model, neutrinos obtain masses $m_\nu \sim h \langle H_u
\rangle/M_N \lesssim 10^{-3} \eV$, where $\langle H_u \rangle$ is the
expectation value of the Higgs field and $M_N$ the sneutrino mass.  In
\cite{lepto} leptogenesis from an sneutrino dominated universe is
considered.  In this model, sneutrino has mass $M_N \sim 10^7 \GeV$
and $h \lesssim 10^{-12}$. The sneutrino can be the curvaton if there
are no constraints from gravitino production.  Moreover, if the
sneutrino mass is generated at some GUT scale, one needs to explain
why $M_N \ll M_{GUT}$ and why non-renormalizable operators are absent
to very high order.

This work was supported by the European Union under the RTN contract
HPRN-CT-2000-00152 Supersymmetry in the Early Universe.


\end{document}